\documentclass[aps,onecolumn,prd,showpacs,nofootinbib]{revtex4}

\usepackage{amsmath}
\usepackage{graphicx}
\usepackage{subfigure}
\usepackage{dcolumn}
\usepackage{bm}
\usepackage{amssymb}
\usepackage{latexsym}

\newcommand{\ie}{\emph{i.e.~}}

\newcommand{\mP}{m_{\mathrm{Pl}}}

\def\spose#1{\hbox to 0pt{#1\hss}}
\def\lta{\mathrel{\spose{\lower 3pt\hbox{$\mathchar"218$}}
     \raise 2.0pt\hbox{$\mathchar"13C$}}}
\def\gta{\mathrel{\spose{\lower 3pt\hbox{$\mathchar"218$}}
     \raise 2.0pt\hbox{$\mathchar"13E$}}}

\newcommand{\mpl}{m_\mathrm{Pl}}
\newcommand{\de}[2]{\kern - #1 em \mathrm{d} #2}

\begin{document}

\title{Curvaton Decay into Baryons, anti-Baryons and Radiation}

\author{Martin Lemoine} \email{lemoine@iap.fr} \affiliation{Institut
d'Astrophysique de Paris, UMR 7095-CNRS, Universit\'e Pierre et Marie
Curie, 98bis boulevard Arago, 75014 Paris, France}

\author{J\'er\^ome Martin} \email{jmartin@iap.fr}
\affiliation{Institut d'Astrophysique de Paris, UMR 7095-CNRS,
Universit\'e Pierre et Marie Curie, 98bis boulevard Arago, 75014
Paris, France}

\author{Gr\'egory Petit} \email{petit@iap.fr}
\affiliation{Institut d'Astrophysique de Paris, UMR 7095-CNRS,
Universit\'e Pierre et Marie Curie, 98bis boulevard Arago, 75014
Paris, France}

\date{\today}

\begin{abstract}
This paper calculates the amount of baryon/radiation isocurvature
fluctuation produced through the decay of a curvaton field. It is shown
in particular that if curvaton decay preserves baryon number and the
curvaton dominates the energy density at the time of decay, the initial
curvaton/radiation isocurvature mode is entirely transfered into a
baryon/radiation isocurvature mode. This situation is opposite to that
previously studied in three fluid models of curvaton decay; this
difference is related to the conservation of the pre-existing baryon
asymmetry and to the efficiency of the annihilation of all
baryon/anti-baryon pairs produced in the decay. We study in detail the
relevant cases in which the curvaton decay preserves or not baryon
number and provide analytical and numerical calculations for each
situation.
\end{abstract}

\maketitle


\section{Introduction}
\label{sec:introduction}

The curvaton
scenario~\cite{Mollerach:1990ue,Lyth:2001nq,Buonanno:2000cp,
Enqvist:2001zp,Lyth:2002my,Moroi:2001ct,Moroi:2002rd,Linde:1996gt,
Lyth:2003ip} is a variant of the inflationary scenario in which the
field driving the accelerated expansion (the inflaton field) is not
necessarily that which produces all of the primordial
fluctuations. Another field (the curvaton field), through its decay, can
seed part of (or even all of) the cosmological perturbations. Therefore,
in the most generic situation, these fluctuations originate from two
different sources and the possibility of having isocurvature modes
arises.

\par

The existence of isocurvature fluctuations would lead to distortions of
the multipole moments of cosmic microwave background anisotropies, as
compared to pure adiabatic modes. It is thus possible to constrain the
fraction of isocurvature modes using high accuracy measurements, and
present-day constraints show that the contribution of isocurvature modes
is sub-dominant, at least if the isocurvature components are considered
separately~\cite{Stompor:1995py,Enqvist:2000hp,Amendola:2001ni,Crotty:2003rz,Gordon:2003hw,Beltran:2004uv,Moodley:2004nz,KurkiSuonio:2004mn,Beltran:2005gr,Bucher:2004an,Seljak:2006bg,Bean:2006qz,Trotta:2006ww}. 
Therefore, the study of the production of isocurvature
fluctuations in the curvaton scenario, although interesting {\it per
se}, can also help us to constrain the free parameters describing the
model.

\par

The phenomenology of the curvaton scenario has been studied in the
literature in a variety of cases, in particular in multi-fluid
configurations (see for instance
Refs.~\cite{Moroi:2001ct,Moroi:2002rd,Lyth:2002my,Lyth:2003ip,
Gupta:2003jc,Ferrer:2004nv,Lemoine:2006sc,Multamaki:2007hv}). The
purpose of the present paper is to apply the formalism developed in
Ref.~\cite{Lemoine:2006sc} to the particular case of a net
baryon/radiation isocurvature mode generated through curvaton decay.
One peculiar feature that will emerge from the present study is the fact
that the curvaton may induce a maximal isocurvature mode even if it
dominates the energy density at the time of its decay, provided its
decay preserves baryon number. This feature stands in sharp contrast
with previous findings which showed that for curvaton decay into
radiation and another fluid such as dark matter, the decay of a
dominating curvaton would erase any pre-existing isocurvature mode. We
study this case in detail and show that this particularity is related to
the conservation of baryon number and to the efficient annihilation of
all ${\rm b\bar b}$ pairs produced in curvaton decay (throughout this
paper, ``b'' stands for a generic baryon, and ${\rm\bar b}$ for its
antiparticle, not to be confused with bottom and anti-bottom quarks).

\par

This paper is organized as follows. In Sec.~\ref{sec:description of the
model}, we describe the model and formulate the equations of motion at
the background and perturbed levels. In Sec.~\ref{sec:results}, we
numerically solve these equations in two cases, namely when the decay is
symmetric in baryons and anti-baryons and when it is asymmetric meaning
that the production of a net baryon number becomes possible. We show
that these two cases correspond to very different
phenomenologies. Finally, In Sec.~\ref{sec:Conclusion}, we discuss and
compare our main results and present our conclusions.

\section{Description of the model}
\label{sec:description of the model}

We consider a model where four fluids are present: baryons (denoted in
what follows with the subscript ``${\rm b}$''), anti-baryons
(``$\bar{\rm b}$''), radiation (``${\rm r}$'') and the curvaton field
(``$\sigma$''). At the fundamental level, the curvaton is {\it a
  priori} a massive scalar field but can effectively be treated as a
pressureless fluid. One assumes that it can decay into radiation,
baryons and anti-baryons. Each of these processes is controlled by a
partial decay width denoted $\Gamma _{\rm \sigma r}$, $\Gamma _{\rm
  \sigma b}$ and $\Gamma _{\rm \sigma \bar{b}}$ respectively. The
curvaton decay occurs when the condition $\Gamma _{\sigma }\sim H$ is
met, where $\Gamma _{\sigma }$ is the total decay width, namely
$\Gamma _{\sigma} =\Gamma _{\rm \sigma r}+\Gamma _{\rm \sigma
  b}+\Gamma_{\rm \sigma \bar{b}}$ and $H$ the Hubble parameter. 

\par

We do not discuss the phenomenology of curvaton to dark matter decay in
the present paper. It is fair to assume that by curvaton decay, dark
matter is effectively decoupled from radiation and
baryons/anti-baryons. Even though the decay of curvaton may induce a
dark matter - radiation isocurvature mode (see
Ref.~\cite{Lemoine:2006sc} for a detailed analysis), or a baryon -
radiation isocurvature mode, both baryonic and dark matter sector will
evolve independently. In this sense, the constraints obtained on dark
matter or baryon isocurvature modes give complementary constraints on
the physics of curvaton decay.

The freeze-out of baryon/anti-baryon annihilations is controlled by
the velocity averaged cross-section
\begin{equation}
\label{eq:crossection}
\left\langle\sigma_{\rm b\bar{b}}v\right\rangle \simeq m_{\pi}^{-2}\, ,
\end{equation}
where $m_{\pi}=135\, \mbox{MeV}$. This relation originates from the
fact that, in the present context, the pion can be viewed as the gauge
boson mediating the strong force. Freeze-out of ${\rm b\bar b}$
annihilations occurs when $\Gamma _{\rm b\vert {\rm f}}\equiv
n_{\rm{\bar{b}}\vert {\rm f}} \left\langle\sigma_{\rm
b\bar{b}}v\right\rangle$ and/or $\Gamma _{\rm \bar{b}\vert {\rm
f}}\equiv n_{\rm{b}\vert {\rm f}} \left\langle\sigma_{\rm
b\bar{b}}v\right\rangle$ are of the order of the expansion rate $H$
which corresponds to a temperature $\sim 20\,$MeV in the absence of
curvaton decay (that is to say, assuming that radiation always dominates
the energy content of the Universe). 

\par

Big-Bang Nucleosynthesis (BBN) puts rather extreme upper bounds on the
amount of energy density injected at temperatures $T\lesssim1\,{\rm
MeV}$~(see Ref.~\cite{Kawasaki:2004qu} for a recent compilation). For
all practical purposes, it suffices to impose that $T_{\rm d}\,\geq
1\,{\rm MeV}$ to satisfy these constraints. Furthermore, the late time
decay of a scalar field at temperatures of order $1-10\,$MeV is a fairly
generic case in the framework of moduli cosmology. As is well known,
such fields generically possess a very large energy density and a very
small decay width $\Gamma_\sigma\, \sim\, m_\sigma^3/m_{\rm Pl}^2$,
hence they decay after big-bang nucleosynthesis if their mass is of the
order of the weak scale. Therefore, in order to reconcile the existence
of such fields with the success of big-bang nucleosynthesis, one has two
choices: either the energy density of these fields at the time of
big-bang nucleosynthesis is very small or their mass is large, leading
to early enough decay. The mass also cannot be arbitrarily large,
otherwise one has to face a hierarchy problem, hence the generic decay
temperature is $1-10\,$MeV. Supersymmetric models with anomaly mediated
supersymmetry breaking provide an explicit realization of particle
physics model building in which the masses of moduli is of the order of
$m_\sigma\,\sim\, 10-100\,$TeV, which leads to decay temperatures of the
moduli/curvaton $T_{\rm d}\, \sim\, 1-10\,$MeV~\cite{Moroi:1999zb}.

\par

The above motivates the present study of the phenomenology of curvaton
decay at temperatures of order $1-10\,$MeV. Out of simplicity, we keep
this temperature fixed to a value $T_{\rm d}\,=\,5.9\,$MeV in our
numerical analysis, which corresponds to a total decay width
$\Gamma_{\sigma}\,=1.6\times10^{-20}\,$MeV. We will argue that the
results obtained remain unchanged if the decay temperature is higher,
in particular if $T_{\rm d}\,\gtrsim \, 20\,$MeV.

\par

At the background level, following the approach of
Ref.~\cite{Lemoine:2006sc}, the above situation can be modelled by the
following set of equations
\begin{eqnarray}
\label{eq:baryons}
\frac{{\rm d}\Omega_{\rm b}}{{\rm d}N} & = & \Omega_{\rm r}\Omega_{\rm b} +
\frac{\Gamma_{\rm \sigma b}}{H}\Omega_{\sigma} -
\frac{3\left\langle\sigma_{\rm b\bar{b}}v\right\rangle \mpl^2}
{8\pi}\frac{H}{m_{\rm b}}\left(\Omega_{\rm b}
\Omega_{\rm \bar{b}}
-\Omega_{\rm b}^{\rm eq}\Omega_{\rm \bar{b}}^{\rm eq}\right)\, ,\\
\label{eq:anti-baryons}
\frac{{\rm d}\Omega_{\rm \bar{b}}}{{\rm d}N} & = &
\Omega_{\rm r}\Omega_{\rm \bar{b}} + \frac{\Gamma_{\rm \sigma
\bar{b}}}{H}\Omega_{\sigma} -
\frac{3\left\langle\sigma_{\rm b\bar{b}}v\right\rangle
\mpl^2}{8\pi}\frac{H}{m_{\rm \bar{b}}}\left(\Omega_{\rm b}\Omega_{
\rm \bar{b}}-\Omega_{\rm b}^{\rm eq}\Omega_{\rm \bar{b}}^{\rm eq}
\right)\, ,\\
\label{eq:radiation}
\frac{{\rm d}\Omega_{\rm r}}{{\rm d}N} & = & 
\left(\Omega_{\rm r}-1\right)\Omega_{\rm r} +
\frac{\Gamma_{\rm \sigma r}}{H}\Omega_{\sigma} +
2\frac{3\left\langle\sigma_{\rm b\bar{b}}v\right\rangle
\mpl^2}{8\pi}\frac{H}{m_{\rm b}}\left(\Omega_{\rm b}
\Omega_{\rm \bar{b}}-\Omega_{\rm b}^{\rm eq}
\Omega_{\rm \bar{b}}^{\rm eq}\right)\, ,\\
\label{eq:curvaton}
\frac{{\rm d}\Omega_{\sigma}}{{\rm d}N} & = &
\Omega_{\rm r}\Omega_{\sigma}
- \frac{\Gamma_{\sigma}}{H}\Omega_{\sigma}\, ,\\
\label{eq:Hubble}
\frac{{\rm d}H}{{\rm d}N} & = &
- \frac{3H}{2}\left(1+\frac{\Omega_{\rm r}}{3}\right)\, .
\end{eqnarray} 
Let us describe these equations in more detail. As usual, the
parameters $\Omega _{(\alpha)}$ are defined as the ratio of the energy
density of the fluid $\alpha $ to the critical energy density, $\Omega
_{(\alpha)}\equiv \rho _{(\alpha )}/\rho _{\rm cr}$. The time variable
is the number of e-folds, $N\equiv \ln a $, where $a$ is the scale
factor. The quantity $\Omega _{\rm b}^{\rm eq}$ is defined by $\Omega
_{\rm b}^{\rm eq}\equiv m_{\rm b}n_{\rm b}^{\rm eq}/\rho _{\rm cr}$,
where $n_{\rm b}^{\rm eq}$ is the particle density at thermal
equilibrium, expressed as:
\begin{eqnarray}
 n_{\rm b}^{\rm eq} & = & 
g\left( \frac{m_{\rm b}T}{2\pi}\right)^{3/2} 
\exp\left(-\frac{m_{\rm b}-\mu_{\rm b}}{T}\right)\, ,
\end{eqnarray} 
with a similar expression for $n_{\rm \bar{b}}^{\rm eq}$. The quantity
$\mu _{\rm b}$ is the chemical potential of the baryons and one has
$\mu _{\rm b}=-\mu _{\rm \bar{b}}$. The temperature $T$ can be
expressed in terms of the variables of the previous system of
equations as:
\begin{equation}
\label{eq:defT}
T=\left(\frac{\pi^2g_*}{ 30}\frac{8\pi}{
  3\mP^2}\right)^{-1/4} H^{1/2}\Omega_{\rm r}^{1/4}\ .
\end{equation}
Note that the above description implicitly assumes that the curvaton
decay products thermalize instantaneously.  This assumption will be
discussed at the end of Section~\ref{sec:results}. One should already
underline that the above ratios $\Gamma_{\sigma\rm b}/\Gamma_{\sigma}$,
$\Gamma_{\sigma\rm \bar b}/\Gamma_{\sigma}$ and $\Gamma_{\sigma \rm
r}/\Gamma_\sigma$ should be understood as characterizing the fraction of
curvaton energy that eventually goes into thermalized ``${\rm b}$'',
``${\rm \bar b}$'' and ``${\rm r}$'', rather than the branching ratios
associated with curvaton decay channels.

\par

For the sake of simplicity, we ignore any temperature dependence of the
function $g_*$ and we take $g_*=10.75$. If we compare with the equations
of motion established in Ref.~\cite{Lemoine:2006sc} in the case where
the curvaton can decay into dark matter $\chi$ (rather than baryons and
anti-baryons), the only difference is that terms like $\Omega _{\chi}^2$
or $\Omega _{\rm \chi, eq}^2$ are replaced by $\Omega_{\rm b}\Omega_{\rm
\bar{b}}$ and $\Omega_{\rm b}^{\rm eq}\Omega_{\rm \bar{b}}^{\rm
eq}$. Notice that, as a consequence, the evolution of the system does
not depend on the chemical potential which cancels out, thanks to the
fact that $\mu _{\rm b}=-\mu _{\rm \bar{b}}$. Finally, there is a factor
$2$ in front of the last term in Eq.~(\ref{eq:radiation}). This factor
originates from the requirement that the total energy density be
conserved.

\par

Let us also discuss how the initial conditions are chosen. Initially, we
start with thermal equilibrium and some baryons/anti-baryons
asymmetry. This implies that
\begin{equation}
\Omega _{\rm b}\Omega _{\rm \bar{b}}=\Omega _{\rm b}^{\rm (eq)}\Omega
_{\rm \bar{b}}^{\rm (eq)}\, ,\quad \Omega _{\rm b}-\Omega _{\rm
\bar{b}}=\delta  \, .
\end{equation}
These two relations lead to
\begin{equation}
\Omega _{\rm b}=\frac{\delta }{2}\left(1+\sqrt{1+\frac{4}{\delta
^2}\Omega _{\rm b}^{\rm (eq)}\Omega _{\rm \bar{b}}^{\rm (eq)}}\right)\,
, \quad \Omega _{\rm \bar{b}}=\frac{\delta
}{2}\left(-1+\sqrt{1+\frac{4}{\delta ^2}\Omega _{\rm b}^{\rm (eq)}\Omega
_{\rm \bar{b}}^{\rm (eq)}}\right)
\end{equation}
Therefore, if the initial values of $\Omega _{\rm b}^{\rm (eq)}$ and
$\delta $ are known, then one can deduce the initial values of $\Omega
_{\rm b}$ and $\Omega_{\rm \bar{b}}$. The quantities $\Omega _{\rm
b}^{\rm (eq)}$ and $\delta $ can be expressed as 
\begin{equation}
\Omega _{\rm b}^{\rm (eq)}=\frac{g}{(2\pi)^{3/2}}\frac{8\pi }{3H^2}
\frac{m^4_{\rm b}}{\mP^2}x^{-3/2}{\rm e}^{-x}\, , \quad \delta =\frac{8\pi
\zeta(3)gm^4_{\rm b}\epsilon _{\rm b} }{3H^2\mP^2\pi ^2x^3}\, ,
\end{equation}
where $x\equiv m_{\rm b}/T$ and where the quantity $\epsilon_{\rm b} $
is defined by
\begin{equation}
\label{eq:defdelta}
\epsilon _{\rm b} \equiv \frac{n_{\rm b}-n_{\rm \bar{b}}}{n_\gamma }
\, .
\end{equation}
The present-day value of the baryon asymmetry is $\epsilon_{\rm
  b}\,\simeq\,5.4\times 10^{-10}$~\cite{Spergel:2006hy}. The initial
  value of $\epsilon_{\rm b}$ well before the freeze-out of ${\rm b\bar
  b}$ annihilations must therefore be tuned in order to reproduce the
  final value after curvaton decay and entropy transfer from $e^+e^-$ to
  the photons. Curvaton decay may dilute any pre-existing asymmetry
  through entropy production or even produce net baryon number if the
  curvaton decay process violates baryon number. In all our calculations
  presented further below, we have tuned this initial asymmetry in order
  to match the observed present-day value.

\par

In order to establish Eq.~(\ref{eq:defdelta}), we have used the fact
that the number of photons is given by $n_{\gamma }=\zeta(3)gT^3/\pi
^2$. It is important to notice that the difference $n_{\rm b}-n_{\rm
  \bar b}$ is normalized with respect to the photon energy density (or
number) and not to the total radiation energy density. In these
formulas, $x=x_{\rm ini}\sim 10$ (for instance) and $m_{\rm b}\sim 0.9
\mbox{GeV}$ are known (or chosen). Moreover, the Hubble parameter and
$\Omega _{\rm r }$ are related through Eq.~(\ref{eq:defT}). Then,
using the fact that the space-like sections are flat, \ie $\Omega
_{\sigma }+\Omega _{\rm r}+\Omega _{\rm b}+\Omega _{\rm \bar{b}}=1$,
and considering the (initial value of) $\Omega _{\sigma ,\rm ini}$ as
a free quantity, one can derive the following expression
\begin{equation}
H^2=\frac{1}{1-\Omega _{\sigma }}\frac{8\pi ^3g_*m^4_{\rm b}}{90 x^4\mP^2}
\left[1\pm \frac{60g x^{5/2}{\rm e}^{-x}}{\pi ^2(2\pi)^{3/2}g_*}
\sqrt{1+\frac{2\zeta^2(3)\epsilon ^2_{\rm b}{\rm e}^{2x}}{\pi x^3}}\right]\, .
\end{equation}
Therefore, for a given value of $\Omega _{\sigma ,\rm ini}$, $H_{\rm
ini}$ can be computed and the other quantities $\Omega _{\rm r,\rm
ini}$, $\Omega _{\rm b, ini}$ and $\Omega _{\rm \bar{b}, ini}$, simply
follow from the above equations.

\par

Let us now consider the perturbations. In order to establish the
gauge-invariant equations of motion, we follow the method of
Ref.~\cite{Lemoine:2006sc}. It consists in formulating the equations in
a covariant way in order to be able to perturb them consistently. One
can write
\begin{eqnarray}
\label{eq:conservation}
\nabla_{\mu}T^{\mu}{}_{\nu(\alpha)} = Q_{\nu(\alpha)}+Y_{\nu(\alpha)}
\end{eqnarray} 	
where $Q^{\mu}=\Gamma T^{\mu\nu}u_{\nu}$ is the curvaton decay term,
$u_{\nu}$ being the four velocity of a fundamental observer, and the
term $Y^{\mu}$ is a phenomenological description of the interaction
term. It reads
\begin{eqnarray}
\label{eq:defY}
Y^{\mu}=\frac{\left\langle\sigma_{\rm b \bar{b}}v\right\rangle}
{m_{\rm b}}\left[
T^{\mu}{}_{\rm \lambda (b)}T^{\lambda\beta}_{\rm (\bar{b})}
-T^{\rm \mu,eq}{}_{\rm \lambda(b)}T^{\rm \lambda\beta,eq}_{\rm 
(\bar{b})}\right]u_{\beta}\, . 
\end{eqnarray}
Of course, a rigorous treatment of the problem would rely on the full
Boltzmann equation but this phenomenological description will be
sufficient for our purpose. In particular, one can check that
Eq.~(\ref{eq:conservation}) exactly reproduces the background
equations~(\ref{eq:baryons})-(\ref{eq:Hubble}). Moreover, it is
straightforward to perturb Eq.~(\ref{eq:conservation}). This leads to
the following system
\begin{eqnarray}
\label{eq:delta b}
\frac{{\rm d}\Delta_{\rm b}}{{\rm d}N} & = & - \frac{\Gamma_{\rm \sigma
    b}}{H}\frac{\Omega_{\sigma}}{\Omega_{\rm b}}
\left(\Delta_{\rm b}-\Delta_{\sigma}
\right) - \frac{3}{2}\left(\Omega_{\sigma}\Delta_{\sigma} +
\Omega_{\rm r}\Delta_{\rm r} + \Omega_{\rm b}\Delta_{\rm b}+
\Omega_{\rm \bar{b}}\Delta_{\rm \bar{b}}\right) - \Phi\left(
3-\frac{\Gamma_{\rm \sigma b}}{H}\frac{\Omega_{\sigma}}{\Omega_{\rm b}}\right)
\nonumber\\ & & -
\frac{3\left\langle\sigma_{\rm b\bar{b}}v\right\rangle
  \mpl^2}{8\pi}\frac{H}{m_{\rm b}\Omega_{\rm b}}\Biggr\{\left[
  \Omega_{\rm b}\Omega_{\rm \bar{b}}\left(
  \Delta_{\rm b}+\Delta_{\rm \bar{b}}\right)
  -2\Omega_{\rm b}^{\rm eq}\Omega_{\rm \bar{b}}^{\rm eq}\Delta^{\rm
  eq}\right]
  +\left(\Phi+\Delta_{\rm b}\right)\left(
  \Omega_{\rm b}\Omega_{\rm \bar{b}}-\Omega_{\rm b}^{\rm eq}
\Omega_{\rm \bar{b}}^{\rm eq}\right)\Biggr\}\, ,
\\
\label{eq:delta barb}
\frac{{\rm d}\Delta_{\rm \bar{b}}}{{\rm d}N} & = & 
- \frac{\Gamma_{\rm \sigma
    \bar{b}}}{H}\frac{\Omega_{\sigma}}{\Omega_{\rm \bar{b}}}
\left(\Delta_{\rm \bar{b}}-\Delta_{\sigma}
\right) - \frac{3}{2}\left(\Omega_{\sigma}\Delta_{\sigma} +
\Omega_{\rm r}\Delta_{\rm r} + \Omega_{\rm b}\Delta_{\rm b}+
\Omega_{\rm \bar{b}}\Delta_{\rm \bar{b}}\right) - \Phi\left(
3-\frac{\Gamma_{\rm \sigma
    \bar{b}}}{H}\frac{\Omega_{\sigma}}{\Omega_{\rm \bar{b}}}\right)
\nonumber\\ & & -
\frac{3\left\langle\sigma_{\rm b\bar{b}}v\right\rangle
  \mpl^2}{8\pi}\frac{H}{m_{\rm \bar{b}}\Omega_{\rm \bar{b}}}
\Biggl\{\left[
  \Omega_{\rm b}\Omega_{\rm \bar{b}}\left(
  \Delta_{\rm b}+\Delta_{\rm \bar{b}}\right)
  -2\Omega_{\rm b}^{\rm eq}\Omega_{\rm \bar{b}}^{\rm eq}
\Delta^{\rm eq}\right]
  +\left(\Phi+\Delta_{\rm \bar{b}}\right)\left(
  \Omega_{\rm b}\Omega_{\rm \bar{b}}-\Omega_{\rm b}^{\rm eq}
\Omega_{\rm \bar{b}}^{\rm eq}\right)\Biggr\}\, ,
\\
\label{eq:delta rad}
\frac{{\rm d}\Delta_{\rm r}}{{\rm d}N} & = & - \frac{\Gamma_{\sigma
    \rm r}}{H}\frac{\Omega_{\sigma}}{\Omega_{\rm r}}
\left(\Delta_{\rm r}-\Delta_{\sigma}
\right) - 2\left(\Omega_{\sigma}\Delta_{\sigma} +
\Omega_{\rm r}\Delta_{\rm r} + \Omega_{\rm b}\Delta_{\rm b} +
\Omega_{\rm \bar{b}}\Delta_{\rm \bar{b}}\right) - \Phi\left(
4-\frac{\Gamma_{\sigma
    \rm r}}{H}\frac{\Omega_{\sigma}}{\Omega_{\rm r}}\right)
\nonumber\\ & & +2
\frac{3\left\langle\sigma_{\rm b\bar{b}}v\right\rangle
  \mpl^2}{8\pi}\frac{H}{m_{\rm b}\Omega_{\rm r}}\Biggl\{\left[
  \Omega_{\rm b}\Omega_{\rm \bar{b}}\left(
  \Delta_{\rm b}+\Delta_{\rm \bar{b}}\right)
  -2\Omega_{\rm b}^{\rm eq}\Omega_{\rm \bar{b}}^{\rm eq}
\Delta^{\rm eq}\right]
  +\left(\Phi+\Delta_{\rm r}\right)\left(
  \Omega_{\rm b}\Omega_{\rm \bar{b}}-\Omega_{\rm b}^{\rm eq}
\Omega_{\rm \bar{b}}^{\rm eq}\right)\Biggr\}\, ,
\\
\label{eq: delta curv}
\frac{{\rm d}\Delta_{\sigma}}{{\rm d}N} & = & - \frac{3}{2}\left(
\Omega_{\sigma}\Delta_{\sigma} + \Omega_{\rm r}\Delta_{\rm r} +
\Omega_{\rm b}\Delta_{\rm b} +
\Omega_{\rm \bar{b}}\Delta_{\rm \bar{b}}\right) - \Phi\left( 3 +
\frac{\Gamma_{\sigma}}{H}\right) \, ,\\
\label {eq: bardeen}
\frac{{\rm d}\Phi}{{\rm d}N} & = & - \Phi-\frac{1}{2}\left(
\Omega_{\sigma}\Delta_{\sigma} + \Omega_{\rm r}\Delta_{\rm r} +
\Omega_{\rm b}\Delta_{\rm b} +
\Omega_{\rm \bar{b}}\Delta_{\rm \bar{b}}\right)\, ,
\end{eqnarray}
where $\Delta_{\alpha}\equiv \Delta\rho_{\alpha}/\rho_{\alpha}$ is the
gauge-invariant density contrast for the fluid $\alpha$. The quantity
$\Delta^{\rm eq}$ is defined by the following expression
\begin{equation}
\Delta^{\rm eq}\equiv \frac14\left(\frac32+x\right)\Delta _{\rm r}\, .
\end{equation}
With this definition, it is easy to see that we deal with a ``closed''
system of equations since $x$ must be viewed as a function of $H$ and
$\Omega_{\rm r}$, see Eq.~(\ref{eq:defT}). Let us now turn to the
discussion of the solutions of the two systems of equations presented in
this section.

\section{Results}
\label{sec:results}

The main parameters that govern the cosmological consequences of
curvaton decay into radiation and baryon/anti-baryons are: {\it (i)}
the time of decay of the curvaton, which is encoded in the total decay
width $\Gamma_\sigma$, {\it (ii)} the respective branching ratios
$\Gamma_{\sigma\rm b}/\Gamma_\sigma$ and $\Gamma_{\rm \sigma\bar
  b}/\Gamma_\sigma$; {\it (iii)} the magnitude of the curvaton energy
density at the time of decay, i.e. $\Omega_\sigma^{<_{\rm d}}$ when
$H=\Gamma_\sigma$. The main parameters are therefore the respective
branching ratios and $\Omega_\sigma^{<_{\rm d}}$. Note that the
branching ratios are constrained by the measured baryon asymmetry
$\epsilon_{\rm b}\,\simeq\, 5.4\times10^{-10}$. In particular the
baryon asymmetry $\epsilon_{\rm b}$ measured immediately after
curvaton decay should be equal to $1.5\times10^{-9}$, in order to
obtain the measured value after the reheating of the photon fluid by
electron/positron annihilations. According to whether $\Gamma_{\rm
  \sigma b}=\Gamma_{\sigma\rm\bar b}$ or not, two possibilities may
arise. In the case of symmetric decay, meaning $\Gamma_{\rm \sigma
  b}=\Gamma_{\sigma\rm\bar b}$, the baryon asymmetry is generated by
some unspecified mechanism acting at a higher energy scale; it is
simply diluted during curvaton decay by the extra entropy brought by
the curvaton. In the case of asymmetric decay, the curvaton
contributes to the net baryon asymmetry. We note that direct
baryogenesis at a low temperature $T\,\sim\,10\,$MeV is very
contrived; we will nevertheless study this case for the sake of
completeness and discuss the robustness of the results for higher
decay temperatures. These two scenarios indeed exhibit different
consequences, as discussed in turn in the following.

\par

We will use the standard definition of the curvature perturbation in
fluid $(\alpha)$~\cite{Lyth:1984gv,Martin:1997zd,Lyth:2003im}:
\begin{equation}
\zeta _{(\alpha)}\,\equiv\, -\Phi -H\frac{\Delta
  \rho_{(\alpha)}}{\dot \rho_{(\alpha)}}\, \simeq -\Phi
  +\frac{\Delta_{(\alpha)}}{3\left[1+\omega_{(\alpha)}\right]}\, ,
\label{eq:zeta}
\end{equation}
where, in order to express $\dot{\rho }_{(\alpha)}$, we have not
considered the interaction term. The corresponding definitions for the
isocurvature modes read
\begin{equation}
S_{\rm b r}\,\equiv\,3\left(\zeta_{\rm b}-\zeta_{\rm r}\right),\quad
S_{\rm\bar{b}r}\,\equiv\,3\left(\zeta_{\rm\bar{b}}-\zeta_{\rm r}\right),
\quad
S_{\sigma\rm r}\,\equiv\,3\left(\zeta_\sigma-\zeta_{\rm r}\right)\ .
\end{equation}
In particular, we will be interested in the transfer of the initial
curvaton/radiation isocurvature perturbation into the final
baryon/radiation isocurvature mode, as expressed by the ratio $S_{\rm
br}^{\rm (f)}/S_{\sigma\rm r}^{\rm (i)}$.  The quantity indexed
with ${\rm (f)}$ [resp. ${\rm (i)}$] is evaluated well after the decay
(resp. well before). In the following, we also express quantities
evaluated immediately before (resp. after) decay with the superscript
$^{<_{\rm d}}$ (resp. $^{>_{\rm d}}$).

\subsection{Symmetric decay}
\label{subsec:symdecay}

In this sub-section, we explore the phenomenology of models in which the
curvaton decays symmetrically into baryons and anti-baryons, \ie
$\Gamma_{\sigma\rm b}=\Gamma_{\sigma\rm\bar b}$. We find that two
situations may arise, according to whether the curvaton dominates the
energy density at its decay, \ie $\Omega_\sigma^{<_{\rm d}}\,\sim\,1$, or
not.

\begin{figure*}
\centerline{
\begin{tabular}{cc}
  \includegraphics[width=0.7\textwidth,clip=true]{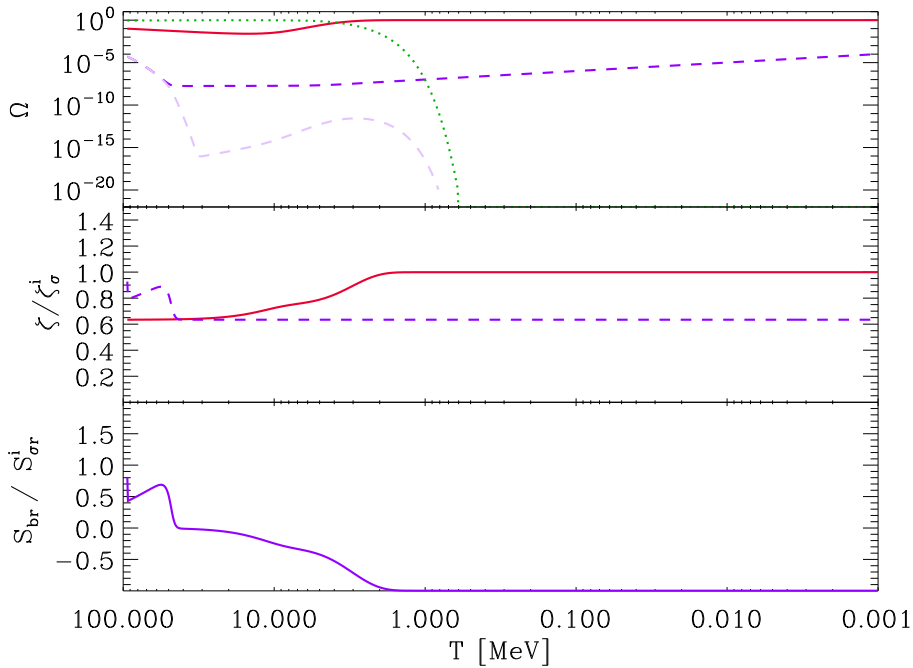}
\end{tabular}}
\caption{Evolution of the background and perturbed quantities in the
  case in which the curvaton decays symmetrically into baryons and
  anti-baryons: $\Gamma_{\sigma\rm b}=\Gamma_{\sigma\rm\bar
  b}=5\times10^{-4}\Gamma_\sigma$. The total decay width is
  $\Gamma_\sigma=1.6\times10^{-20}\,$MeV, and $\Omega_\sigma^{\rm
  (i)}=0.9$ at a temperature $T=\,94\,$MeV. The top panel shows the
  evolution of $\Omega_{\rm r}$ (solid red line), $\Omega_\sigma$
  (dotted green line), $\Omega_{\rm b}$ (upper dashed dark blue line)
  and $\Omega_{\rm\bar b}$ (lower dashed light blue line). The middle
  panel shows the evolution of $\zeta_{\rm r}$ (solid red line) and
  $\zeta_{\rm b}$ (dashed blue line). The bottom panel shows the
  evolution of the isocurvature transfer coefficient $S_{\rm b
  r}/S_{\sigma\rm r}^{\rm (i)}$. The transfer of isocurvature
  perturbation is maximal as the radiation but not the baryon fluid
  inherits the curvaton perturbation.}  \label{fig:1}
\end{figure*}

Consider first the case in which the curvaton dominates the energy
density at decay, $\Omega_\sigma^{<_{\rm d}}\,\sim\,1$. We find that the
transfer coefficient of the isocurvature mode is maximal, as exemplified
for instance in Fig.~\ref{fig:1}. The top panel of this figure shows
the evolution of the background energy density in radiation (solid red
line), curvaton (dotted green line), in baryons (top dashed dark blue
line) and in anti-baryons (bottom dashed light blue line). The middle
panel shows the evolution of the individual $\zeta_{(\alpha)}$
quantities, and the bottom panel the transfer of the isocurvature
fluctuation. This latter shows clearly the emergence of a net
baryon-radiation isocurvature fluctuation after the decay of the
curvaton. The middle panel also reveals that in this case, the
radiation fluid inherits the curvaton fluctuations (since
$\zeta_\gamma/\zeta_\sigma^{\rm (i)}\rightarrow 1$ at $T\ll T_{\rm d}$),
while the baryon fluid remains unaffected.

\par

This result is quite different from a ``standard'' scenario of
curvaton decay into radiation and dark matter, in which the domination
of the curvaton at the time of decay ensures that only adiabatic modes
subsist, as all fluids have inherited the same curvaton
perturbations. This difference can be related to the annihilation of
all ${\rm b\bar b}$ pairs produced by the curvaton, which effectively
reduces to zero the net energy transfer of the curvaton to the baryon
fluid. In order to put this statement on quantitative footing, it is
useful to evaluate the ratio of the annihilation rates of baryons and
anti-baryons to the expansion rate immediately after curvaton decay
($H=\Gamma_\sigma$):
\begin{eqnarray}
\Upsilon_{\rm b}&\,=\,&{n_{\rm\bar b}^{>_{\rm d}}\langle\sigma_{\rm b\bar b}v\rangle
  \over \Gamma_\sigma}\,\simeq\, 4.6\times10^{16}\,\left({T_{\rm
    d}\over 10\,{\rm MeV}}\right)^2\,\Omega_{\rm\bar b}^{>_{\rm
    d}}\ ,\nonumber\\
\Upsilon_{\rm \bar b}&\,=\,&{n_{\rm b}^{>_{\rm d}}\langle\sigma_{\rm b\bar b}v\rangle
  \over \Gamma_\sigma}\,\simeq\, 4.6\times10^{16}\,\left({T_{\rm
    d}\over 10\,{\rm MeV}}\right)^2\,\Omega_{\rm b}^{>_{\rm
    d}}\ .
\end{eqnarray}
The first equation gives the ratio $\Upsilon_{\rm b}$ of the
annihilation rate of baryons to the expansion rate, while the second
gives the corresponding ratio $\Upsilon_{\rm\bar b}$ of the
annihilation rate of anti-baryons to the expansion rate. Considering
$\Upsilon_{\rm b}$, the above formula shows that if $\Omega_{\rm\bar
  b}$ exceeds $\sim 10^{-16}$, annihilations are effective. In the
absence of curvaton, the freeze-out of ${\rm b\bar b}$ annihilations occurs
as the abundance of anti-baryons is reduced to below this threshold. In
the presence of a curvaton however, the decay of this field will
regenerate the annihilations provided the amount of curvaton produced
anti-baryons is sufficient, i.e. $\Omega_{\rm\bar b}^{>_{\rm
    d}}\,\sim\, \Omega_{\sigma}^{<_{\rm d}}\,\Gamma_{\sigma\rm\bar
  b}/\Gamma_{\sigma} \,\sim\, \Gamma_{\sigma\rm\bar b}/\Gamma_{\sigma}
\gtrsim 10^{-16}$. Then all pairs of baryon/anti-baryon produced by
curvaton decay will annihilate. Of course, if the branching ratio
$\Gamma_{\sigma\rm\bar b}/\Gamma_{\sigma}\lesssim 10^{-16}$, the
regeneration of annihilations will not take place, but the curvaton
will not exert any influence on the pre-existing baryon fluid either.

\par

In fact, the behaviors of the different quantities plotted in
Fig.~\ref{fig:1} can be understood in more detail along the following
lines. Consider the variables associated to net baryon number, in
particular $\Omega_{\rm b}-\Omega_{\rm\bar b}$. Its equation of motion
reduces to:
\begin{equation}
{\mathrm{d}\over\mathrm{d}N}\,\left(\Omega_{\rm b} - \Omega_{\rm\bar
  b}\right)\, =\, \Omega_{\rm r}\left(\Omega_{\rm b} - \Omega_{\rm\bar
  b}\right)\ .\label{eq:netb}
\end{equation}
This composite fluid is isolated, as neither annihilation nor curvaton
decay violates baryon number. Therefore, the curvature perturbation
associated to this composite fluid is conserved, as predicted in
Ref.~\cite{Lyth:2002my}.  Furthermore, one has $\Omega_{\rm\bar
  b}\,\ll\,\Omega_{\rm b}$ before curvaton decay and after
annihilation freeze-out, so that this fluid of ``net baryon number''
approximately corresponds to the baryon fluid.  As annihilations of
${\rm b\bar b}$ pairs produced by curvaton decay is efficient, the
above inequality remains valid after curvaton decay, hence ``net
baryon number'' remains a good approximation for the baryon fluid. All
in all, the above indicates that the curvature perturbation of the
baryon fluid should remain conserved if curvaton decay preserves
baryon number and if annihilations of ${\rm b\bar b}$ pairs are
efficient.

\par

This implies that the above results remain unchanged if the decay
temperature $T_{\rm d}\,\gtrsim\,20\,$MeV. The theorem of
Ref.~\cite{Weinberg:2004kf} stipulates that the isocurvature mode
between two fluids sharing thermal equilibrium are erased on a small
timescale, unless there exists a conserved charge. In the present case,
baryon number is conserved, or more precisely net baryon number does not
couple to radiation, hence the above theorem does not
apply. Consequently, once the isocurvature mode is produced, it remains
conserved unless baryon number violating processes take place. In other
words, one can extrapolate the above results to temperatures at least as
high as the electroweak scale. Let us also remark that, if baryons are
relativistic (at temperatures above the QCD scale), the above equations
are slightly modified, but the above results remain unmodified.

\par

One a more formal level, one can follow the evolution of the different
variables as follows. Neglecting $\Omega_{\rm\bar b}$ in front of
$\Omega_{\rm b}$ in Eq.~(\ref{eq:netb}) above indicates that
$\Omega_{\rm b}$ scales as $a$ when $\Omega_{\rm r}\,\sim\,1$ (\ie
after curvaton decay), while $\Omega_{\rm b}$ remains approximately
constant when $\Omega_{\rm r}\,\ll\,1$. These trends are observed in
Fig.~\ref{fig:1}.

\par

The behavior of $\Omega_{\rm\bar b}$ is less trivial to obtain (but its
cosmological relevance is also much less). One can approximate
Eq.~(\ref{eq:anti-baryons}) with the following, after curvaton decay:
\begin{equation}
{\mathrm{d}\Omega_{\rm\bar b}\over\mathrm{d}N}\,\simeq\,-{3m_{\rm
    Pl}^2\over 8\pi}{\langle\sigma_{\rm b\bar b}v\rangle H\over
  m_{\rm b}}\Omega_{\rm b}\Omega_{\rm\bar b}\ ,
\end{equation}
where the term $\Omega_{\rm r}\Omega_{\rm\bar b}$ has been neglected as
the annihilations are dominant. Using the fact that $H\propto a^{-2}$
and $\Omega_{\rm b}\propto a$ after curvaton decay, one derives the
following late time value of $\Omega_{\rm\bar b}$:
\begin{equation}
\Omega_{\rm\bar b}^{\rm (f)}\,\simeq\, \Omega_{\rm\bar b}^{>_{\rm
    d}}\,{\rm e}^{-\Upsilon_{\rm\bar b}^{>_{\rm d}}}\ .
\end{equation} 
Since $\Upsilon_{\rm\bar b}^{>_{\rm d}}$ takes enormous values of order
$10^9$, the annihilations regenerated by curvaton decay essentially
erase all trace of anti-baryons and the corresponding plateau cannot be
observed in Fig.~\ref{fig:1} because it is too small.

\par

Let us now turn to the perturbations and assume that curvaton decay is
instantaneous. If the curvaton dominates the energy density before
decay, and transfers its energy to radiation, then:
\begin{equation}
\zeta_{\rm r}^{\rm (f)}\,\simeq\,\zeta_\sigma^{\rm (i)}\ .\label{eq:evolzg}
\end{equation}
This relation can be obtained through standard methods and corresponds
to the conservation of the total curvature perturbation throughout
curvaton decay. Similarly, one can build the variable associated to the
perturbation of net baryon number, $\Omega_{\rm b}\Delta_{\rm
b}-\Omega_{\rm\bar b}\Delta_{\rm\bar b}$, which for all practical
purposes, can be approximated by $\Omega_{\rm b}\Delta_{\rm b}$. The
equation of motion for this quantity reads:
\begin{equation}
{\mathrm{d}\over\mathrm{d}N}\left(\Omega_{\rm b}\Delta_{\rm
  b}-\Omega_{\rm\bar b}\Delta_{\rm\bar
  b}\right)\,=\,3{\mathrm{d}\Phi\over\mathrm{d}N}\left(\Omega_{\rm b}
- \Omega_{\rm\bar b}\right)\,+\,\Omega_{\rm r} \left(\Omega_{\rm
  b}\Delta_{\rm b}-\Omega_{\rm\bar b}\Delta_{\rm\bar b}\right)\ .
\end{equation}
Since $\Phi$ is conserved both before and after curvaton decay, the
first term on the r.h.s. can be neglected, and $\Omega_{\rm
b}\Delta_{\rm b}-\Omega_{\rm\bar b}\Delta_{\rm\bar b}$ is approximately
conserved when $\Omega_{\rm r}\,\ll \,1$. Approximating $\Omega_{\rm
b}\Delta_{\rm b}-\Omega_{\rm\bar b}\Delta_{\rm\bar b}$ with $\Omega_{\rm
b}\Delta_{\rm b}$, this implies that $\Delta_{\rm b}$ is approximately
conserved, since $\Omega _{\rm b}$ is constant in this case (see before)
and, hence, that $\zeta_{\rm b}$ is also conserved. At late times, after
curvaton decay, $\Omega_{\rm r}\sim1$ implies that $\Omega_{\rm
b}\Delta_{\rm b}$ scales as $a$, hence that $\Delta_{\rm b}$ (and
therefore $\zeta_{\rm b}$) is again approximately constant because
$\Omega _{\rm b}\propto a$. One thus finds that:
\begin{equation}
\zeta_{\rm b}^{\rm (f)}\,\simeq\, \zeta_{\rm b}^{\rm (i)}\ .\label{eq:evolzb}
\end{equation}
As mentioned above, this property can be traced back to the fact that
net baryon number behaves in the present case as an isolated fluid,
hence its curvature perturbation is a conserved quantity.  Finally, one
derives from Eq.~(\ref{eq:evolzg}) and (\ref{eq:evolzb}) above the
transfer of isocurvature perturbation:
\begin{equation}
S_{\rm b\rm r}^{\rm (f)}\,\simeq\, - S_{\sigma\rm r}^{\rm (i)}\quad\quad
(\Omega_\sigma^{<_{\rm d}}\,\simeq\,1)\ .
\end{equation}
These results match the numerical evolution observed in Fig.~\ref{fig:1}.

\begin{figure*}
\centerline{
\begin{tabular}{cc}
  \includegraphics[width=0.7\textwidth,clip=true]{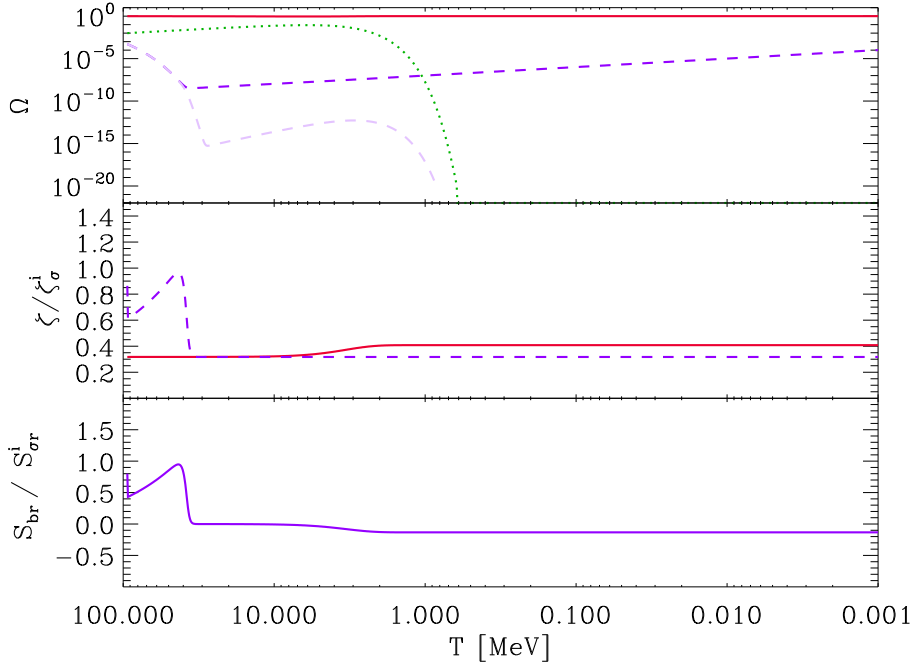}
\end{tabular}}
\caption{Same as figure \ref{fig:1} except that $\Omega_\sigma^{\rm
    (i)}=0.01$, which corresponds to $\Omega_\sigma^{<_{\rm
      d}}\,=\,0.09$. Essentially no isocurvature fluctuation is
  produced as neither the baryon nor the radiation fluctuations have
  been affected by curvaton decay.}
\label{fig:2}
\end{figure*}

Obviously, the above discussion suggests that $S_{\rm b\rm r}^{\rm
  (f)}\rightarrow 0$ as $\Omega_\sigma^{<_{\rm d}}\rightarrow 0$ since
  the net baryon number must remain unaffected, while a decreasing
  curvaton energy density at the time of decay implies that a lesser
  amount of radiation is produced during the decay. In more detail, one
  should obtain
\begin{equation}
S_{\rm b r}^{\rm (f)}\,\simeq\, -\Omega_{\sigma}^{<_{\rm
    d}}S_{\sigma\rm r}^{\rm (i)}\ ,\label{eq:transsym}
\end{equation}
since
\begin{equation}
\zeta_{\rm r}^{\rm (f)}\,\simeq\,\left(1-\Omega_{\sigma}^{<_{\rm
    d}}\right)\zeta_{\rm r}^{\rm (i)}\,+\, \Omega_{\sigma}^{<_{\rm
    d}}\zeta_{\sigma }^{\rm (i)}\ .\label{eq:zetr}
\end{equation}
This trend is confirmed in Fig.~\ref{fig:2} which provides an example
with $\Omega_{\sigma}^{\rm (i)}=0.01$, corresponding to
$\Omega_\sigma^{<_{\rm d}}\,\simeq\,0.09$ at decay ($T_{\rm
  d}\,\simeq\,5.9\,$MeV). The final transfer coefficient is of order
$-\Omega_\sigma^{<_{\rm d}}$ as expected.

\subsection{Asymmetric decay}
\label{subsec:asymdecay}

If the inflaton can decay asymmetrically, $\Gamma_{\sigma\rm b}\,
\neq\,\Gamma_{\sigma\rm \bar b}$, the phenomenology is different, as all
``${\rm b}$'' and ``$\bar {\rm b}$'' produced by the curvaton will not
be able to annihilate with each other. In particular, the production of
net baryon number during curvaton decay comes with the transfer of the
curvaton perturbations to the baryon fluid. As already mentioned, known
models of baryogenesis produce baryon number at a much higher scale than
$1-10\,$MeV. We nevertheless discuss this asymmetric case for the sake
of completeness and because it provides useful insights into curvaton
cosmology. Moreover, as we have argued in the previous section, the
present results can be extrapolated to a higher decay temperature,
possibly as high as the electroweak scale.

\par

In the present case, one may expect cosmological consequences opposite
to those found in the case of symmetric decay: if the curvaton
dominates the energy density of the Universe shortly before decaying,
and produces during its decay most of the baryon number, both baryon
and radiation fluid will inherit its perturbations, hence there should
be no final baryon/radiation isocurvature mode. On the contrary, if
the curvaton energy density is small compared to the radiation energy
density shortly before decay, but the curvaton still produces most of
the baryon number, a maximal isocurvature mode between baryon and
radiation should be produced.

\par

\begin{figure*}
\centerline{
\begin{tabular}{cc}
  \includegraphics[width=0.7\textwidth,clip=true]{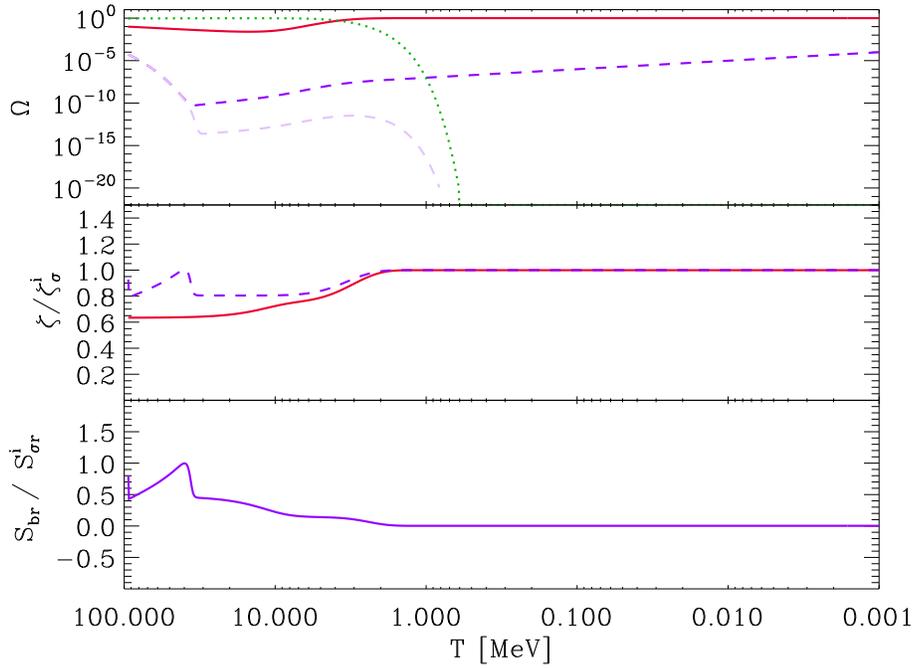}
\end{tabular}}
\caption{Same as figure \ref{fig:1} except that curvaton decay now
  violates baryon number, with $\Gamma_{\sigma\rm
    b}-\Gamma_{\rm \sigma\bar b}=1.7\times10^{-8}\Gamma_\sigma$. The final
  baryon number matches the observed value, for an initial asymmetry
  $\epsilon_{\rm b}=10^{-13}$. Other quantities remain unchanged,
  in particular $\Omega_\sigma^{\rm (i)}=0.9$ and
  $\Gamma_{\sigma}=1.6\times10^{-20}\,$MeV.  Essentially no
  baryon/radiation isocurvature fluctuation results, since the baryon
  and the radiation fluctuations have been similarly affected by
  curvaton decay.}
\label{fig:3}
\end{figure*}

\begin{figure*}
\centerline{
\begin{tabular}{cc}
  \includegraphics[width=0.7\textwidth,clip=true]{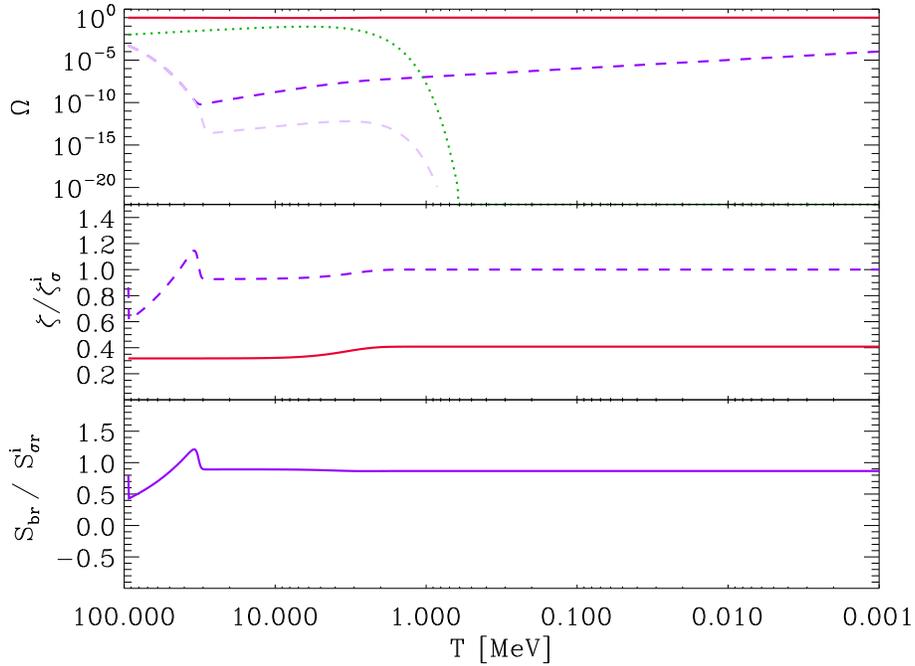}
\end{tabular}}
\caption{Same as figure \ref{fig:3} for baryon violating curvaton
  decay, except that $\Omega_\sigma^{\rm (i)}=0.01$, which corresponds
  to $\Omega_\sigma^{<_{\rm d}}\,=\,0.09$. The baryon violating decay
  width is such that the final baryon number produced matches the
  observed value; this corresponds to $\Delta B_{\rm b\bar
    b}=1.3\times10^{-7}$ for an initial asymmetry $\epsilon_{\rm
    b}=10^{-13}$. A large isocurvature fluctuation is produced as the
  baryon (but not the radiation) fluctuations have been affected by
  curvaton decay.}
\label{fig:4}
\end{figure*}

These trends are confirmed by the numerical computations, as shown in
Figs.~\ref{fig:3} and ~\ref{fig:4}. The first figure,
Fig.~\ref{fig:3}, corresponds to the same value of
$\Omega_{\sigma}^{\rm (i)}$ as in Fig.~\ref{fig:1}, but with an
asymmetric decay width $(\Gamma_{\sigma\rm b}-\Gamma_{\sigma\rm\bar
  b})/\Gamma_\sigma=2\times10^{-8}$. In what follows, we will use the
short-hand notation:
\begin{equation}
\Delta B_{\rm b\bar b}\,\equiv\, {\Gamma_{\sigma\rm b}-\Gamma_{\sigma\rm\bar b}\over
  \Gamma_\sigma}\ .
\end{equation}
Assuming that the initial baryon asymmetry vanishes and that curvaton
decay is instantaneous, one can obtain an order of magnitude of the
decay asymmetry needed to reach the observed value of $\epsilon_{\rm
  b}$ as follows:
\begin{equation}
\Omega_{\rm b}^{>_{\rm d}}-\Omega_{\rm\bar b}^{>_{\rm
    d}}\,\simeq\,\Delta B_{\rm b\bar b}\Omega_\sigma^{<_{\rm d}}\ ,
\end{equation}
which implies:
\begin{equation}
\epsilon_{\rm b}\,\simeq\,7.3\times
10^{-2}\,\left({\Gamma_\sigma\over 10^{-20}\,{\rm
    MeV}}\right)^{1/2}\,\Delta B_{\rm b\bar b}\,\Omega_\sigma^{<_{\rm d}}\ .
\end{equation}
Numerical calculations differ from this simple estimate by a factor of
order unity.

In order to understand these results, it is instructive to express the
time evolution of the baryon asymmetry using the system of
Eqs.~(\ref{eq:baryons}), (\ref{eq:anti-baryons}), (\ref{eq:radiation})
and (\ref{eq:Hubble}).  The baryon asymmetry can indeed be written as:
\begin{equation}
\epsilon_{\rm b}\,\equiv\, 
{\pi^4\over 60\zeta(3)}\left({45\over
  4\pi^3}\right)^{1/4}g_*^{3/4}\left({m_{\rm Pl}\over
  m_{\rm b}}\right)^{1/2}
\left({H\over m_{\rm b}}\right)^{1/2}{\Omega_{\rm b} -
  \Omega_{\rm\bar b}\over \Omega_{\rm r}^{3/4}}
\simeq 2.24\times 10^{10}\left({H\over m_{\rm b}}\right)^{1/2}
{\Omega_{\rm b} -
  \Omega_{\rm\bar b}\over \Omega_{\rm r}^{3/4}}\, .
\end{equation}
Hence the time evolution of the baryon asymmetry is governed by the
following equation:
\begin{equation}
{1\over \epsilon_{\rm b}}{\mathrm{d}\epsilon_{\rm b}\over
  \mathrm{d}N}\,\simeq\, {\Omega_\sigma\over \Omega_{\rm b} -
  \Omega_{\rm\bar b}}{\Gamma_{\sigma\rm b}-\Gamma_{\sigma\rm\bar
    b}\over H} \,-\, {3\over 4} {\Omega_\sigma\over
  \Omega_{\rm r}}{\Gamma_{\sigma\rm r}\over H}\ .
\label{eq:epsilonb}
\end{equation}
In order to obtain the above equation, we have neglected the
baryon/anti-baryon annihilation term in the equation for
$\Omega_{\rm r}$ [Eq.~(\ref{eq:radiation})], which is justified insofar
as the amount of radiation produced in baryon/anti-baryon
annihilations is negligible at or after freeze-out. 

\par 

The above equation is interesting because it shows how the baryon number
can be modified: either through baryon number violating curvaton decay
(first term on the r.h.s), or through dilution due to entropy production
(second term on the r.h.s). It also provides an estimate of the
conditions under which the initial curvaton/radiation isocurvature mode
is efficiently transfered to the baryon/radiation mixture. Such an
efficient transfer can indeed be achieved if $\vert\Delta\epsilon_{\rm
b}/\epsilon_{\rm b}\vert\,\approx\,1$ at curvaton decay, without
significant production of radiation by the curvaton. The latter
condition amounts to negligible entropy production, or what is
equivalent, to assuming that the second term on the r.h.s of
Eq.~(\ref{eq:epsilonb}) is negligible compared to unity. The former
condition then implies that the first term on the r.h.s of
Eq.~(\ref{eq:epsilonb}) is larger than unity. All in all, efficient
transfer of the isocurvature mode occurs if:
\begin{equation}
\Delta B_{\rm b\bar b}\Omega_\sigma^{<_{\rm d}}\,
\gtrsim\, \Omega_{\rm b}^{<_{\rm
    d}}-\Omega_{\rm\bar b}^{<_{\rm d}}\ ,\quad\quad
{\Gamma_{\sigma\rm r}\over\Gamma_\sigma}\Omega_\sigma^{<_{\rm
    d}}\,\lesssim\,\Omega_{\rm r}^{<_{\rm d}}\ .\label{eq:cond}
\end{equation}
It is interesting to remark that this situation is very similar to that
encountered for curvaton decay in a three-fluid model incorporating
radiation and dark matter. Borrowing from the method of
Refs.~\cite{Gupta:2003jc,Lemoine:2006sc}, it is possible to express the
final baryon/radiation isocurvature fluctuation in terms of the initial
curvaton and radiation curvature modes, as follows. One first constructs
a composite fluid that has the property of being isolated, with energy
density:
\begin{equation}
\rho_{\rm comp}\,=\,\rho_{\rm b}\, -\, \rho_{\rm \bar b}\,+\, \Delta
B_{\rm b\bar b}\rho_\sigma\ .
\end{equation}
Notice that this construction is possible because each component of the
composite fluid is pressureless. Its curvature perturbation, which is
conserved by construction, is:
\begin{equation}
\zeta_{\rm comp}\,=\, {\Omega_{\rm b}\over \Omega_{\rm b} -
  \Omega_{\rm \bar b} + \Delta B_{\rm b\bar b}\Omega_\sigma}\zeta_{\rm
  b} -
{\Omega_{\rm\bar b}\over \Omega_{\rm b} -
  \Omega_{\rm \bar b} + \Delta B_{\rm b\bar b}\Omega_\sigma}\zeta_{\rm
  \bar b} +
{\Delta B_{\rm b\bar b}\Omega_\sigma\over \Omega_{\rm b} -
  \Omega_{\rm \bar b} + \Delta B_{\rm b\bar
    b}\Omega_\sigma}\zeta_{\sigma}\ .
\end{equation}
Then, assuming that curvaton decay is instantaneous, one can match the
value of $\zeta_{\rm comp}$ after decay to that before decay, which
gives:
\begin{equation}
\zeta_{\rm b}^{>_{\rm d}}\,\approx\,\zeta_{\rm comp}^{>_{\rm d}}\,=\,
\zeta_{\rm comp}^{<_{\rm d}}\ .
\end{equation}
In order to obtain the first equality, we have used the fact that
$\Omega_{\rm \bar b}^{>_{\rm d}}\,\ll\,\Omega_{\rm b}^{>_{\rm d}}$ as a
result of the efficient annihilation of ${\rm b\bar b}$ pairs after
curvaton decay. Although the quantity $\zeta_{\rm comp}^{<_{\rm d}}$ is
evaluated here immediately before decay, it can be evaluated at any
initial time, since it is conserved.

\par

The radiation perturbation is given by Eq.~(\ref{eq:zetr}), hence the
final baryon/radiation isocurvature perturbation can be written as:
\begin{equation}
S_{\rm br}^{\rm (f)}\,=\,\left[{\Delta B_{\rm b\bar
    b}\Omega_\sigma^{\rm (i)}\over \Omega_{\rm b}^{\rm (i)} -
  \Omega_{\rm\bar b}^{\rm (i)} + \Delta B_{\rm b\bar
    b}\Omega_{\sigma}^{\rm (i)}} - \Omega_\sigma^{<_{\rm
    d}}\right]S_{\sigma\rm r}^{\rm (i)}\ ,
\end{equation}
where we used the fact that $S_{\rm b r}^{\rm (i)}=S_{\rm \bar{b}
r}^{\rm (i)}=0$. As expected, the isocurvature transfer vanishes as
$\Omega_\sigma^{<_{\rm d}}\rightarrow 0$ (since this also implies
$\Omega_\sigma^{\rm (i)}\rightarrow 0$). When $\Omega_\sigma^{<_{\rm
d}}\rightarrow 1$, one can see that the first term in the bracket on the
r.h.s. of the above equation also tends to one, and therefore the
transfer coefficient of the isocurvature mode also vanishes.  The
initial isocurvature fraction is transfered efficiently only if the
conditions expressed in Eq.~(\ref{eq:cond}) are fulfilled. Note also
that in the limit $\Delta B_{\rm b\bar b}\rightarrow 0$, one recovers
the result of Section~\ref{subsec:symdecay} presented in
Eq.~(\ref{eq:transsym}).

\par

Finally, a last point is to be made concerning the assumption of
instantaneous thermalization of the curvaton decay products. If the
center of mass energy $\sqrt{s}\,\sim\,(EE_{\rm th})^{1/2}$ for an
interaction between a high energy particle of energy $E$ and a
thermalized particle of energy $E_{\rm th}$ is well above the QCD scale,
then the ratio of rates of thermalization processes to ${\rm b\bar b}$
producing ones is of the order of $(\alpha_{\rm em}/\alpha_{\rm
s})^2\,\ll\,1$. It is even less if $\sqrt{s}$ is smaller than the QCD
scale. Therefore the above approximation is not strictly speaking
justified. However the neglect of these additional interactions would
not modify our conclusions, for the following reason.

\par

The only effect that could modify our conclusions is if one fluid
(either radiation or baryon) were ``contaminated'' by the other fluid
(respectively baryon or radiation) through the interaction of high
energy particles produced through curvaton decay with thermalized
particles. One typical example is given by the transfer of energy from
the photon to the baryon fluid through $\gamma+\gamma_{\rm
  th}\,\rightarrow\,{\rm b}+{\rm \bar b}$, where $\gamma$ stands for a
high energy photon. However, net baryon number does not couple to
radiation, hence transfers of energy between these two fluids cannot
take place after curvaton decay (provided this latter occurs after any
baryogenesis event).  

\par

Hence all the conclusions remain unaffected by these processes that
occur between curvaton decay and thermalization. It is important to
stress, however, that $\Gamma_{\sigma\rm r}/\Gamma_\sigma$,
$\Gamma_{\sigma\rm b}/\Gamma_\sigma$ and $\Gamma_{\sigma\rm\bar
b}/\Gamma_\sigma$ should not be interpreted strictly speaking as the
branching ratios of curvaton decay into radiation, baryons or
anti-baryons, but rather as the fraction of curvaton energy eventually
transfered into these fluids after all thermalization processes have
occured.

\section{Conclusions}
\label{sec:Conclusion}

In this section, we recap our main results. We have studied the
production of isocurvature perturbations in the curvaton scenario
where the curvaton field can decay into radiation, baryons and
anti-baryons. Two different cases have been considered. The first one
is the symmetric case in which the curvaton/baryon decay width equals
the curvaton/anti-baryon one, i.e. curvaton decay preserves baryon
number.  We have found that if the curvaton dominates the energy
density before decay, then a baryon/radiation isocurvature mode can be
produced. In the opposite situation in which the curvaton contributes
negligibly to the total energy density immediately before decaying,
the isocurvature mode vanishes. This result is opposite to the
standard prediction of the simplest curvaton scenario in which any
pre-existing isocurvature mode is erased by curvaton decay if this
latter dominates the energy density at the time of decay. This
difference can be traced back to the conservation of baryon number and
to the annihilation of all ${\rm b\bar b}$ pairs produced during
curvaton decay.  

\par

One noteworthy consequence of the above is to forbid the liberation of a
significant amount of entropy by a late decaying scalar field at
temperatures below any baryon violating processes, such as is often
invoked for the dilution of unwanted relics.

\par 

Another consequence of the above is that a baryon-radiation
isocurvature mode $S_{\rm b r}$ cannot co-exist with a (WIMP) dark
matter - radiation isocurvature mode $S_{\chi\rm r}$, since the
conditions to produce these modes are opposite to one another. Since
$S_{\chi\rm b}=S_{\chi\rm r}-S_{\rm b r}$, the existence of a
baryon-dark matter isocurvature mode appears generic in this case
(unless $\Omega_{\sigma}^{<_{\rm d}}$ is so small at the time of decay
that the curvaton exerts essentially no influence on dark matter and
baryon perturbations).

\par

The asymmetric decay presents a different phenomenology. Since the
curvaton decay does not produce the same number of baryons and
anti-baryons, the annihilations cannot suppress all the baryonic decay
product and, as a consequence, when the curvaton dominates at decay,
the isocurvature perturbations are erased. In this case, most or all
of the baryon and radiation fluctuations indeed originate from the
curvaton. If the curvaton contribution to the energy density is
smaller than unity at the time of decay, then radiation cannot be
affected substantially, while the baryon fluid may be strongly
affected; this situation results in a large baryon/radiation
isocurvature fluctuation. In some sense, this case appears similar to
the case of curvaton to dark matter decay studied in
Ref.~\cite{Lemoine:2006sc}. Contrary to the previous symmetric case,
non vanishing $S_{\rm b r}$ and $S_{\chi\rm r}$ can co-exist. We note
however, that baryogenesis at low scales (below the electroweak phase
transition) is rather contrived.

\par

On more general grounds, the study presented in this article
exemplifies how scenarios where scalar fields can decay at late times
can be constrained not only at the background level, as it is usually
done, but also by investigating the consequences at the perturbed
level. It is clear that, if this type of information is taken into
account, one can hope to improve our understanding of the feasibility
of such theories. We hope to return to this question in future
publications.


\bibliography{references}

\begin{thebibliography}{33}
\expandafter\ifx\csname natexlab\endcsname\relax\def\natexlab#1{#1}\fi
\expandafter\ifx\csname bibnamefont\endcsname\relax
  \def\bibnamefont#1{#1}\fi
\expandafter\ifx\csname bibfnamefont\endcsname\relax
  \def\bibfnamefont#1{#1}\fi
\expandafter\ifx\csname citenamefont\endcsname\relax
  \def\citenamefont#1{#1}\fi
\expandafter\ifx\csname url\endcsname\relax
  \def\url#1{\texttt{#1}}\fi
\expandafter\ifx\csname urlprefix\endcsname\relax\def\urlprefix{URL }\fi
\providecommand{\bibinfo}[2]{#2}
\providecommand{\eprint}[2][]{\url{#2}}

\bibitem[{\citenamefont{Mollerach}(1990)}]{Mollerach:1990ue}
\bibinfo{author}{\bibfnamefont{S.}~\bibnamefont{Mollerach}},
  \bibinfo{journal}{Phys. Lett.} \textbf{\bibinfo{volume}{B242}},
  \bibinfo{pages}{158} (\bibinfo{year}{1990}).

\bibitem[{\citenamefont{Lyth and Wands}(2002)}]{Lyth:2001nq}
\bibinfo{author}{\bibfnamefont{D.~H.} \bibnamefont{Lyth}} \bibnamefont{and}
  \bibinfo{author}{\bibfnamefont{D.}~\bibnamefont{Wands}},
  \bibinfo{journal}{Phys. Lett.} \textbf{\bibinfo{volume}{B524}},
  \bibinfo{pages}{5} (\bibinfo{year}{2002}), \eprint{hep-ph/0110002}.

\bibitem[{\citenamefont{Buonanno et~al.}(2000)\citenamefont{Buonanno, Lemoine,
  and Olive}}]{Buonanno:2000cp}
\bibinfo{author}{\bibfnamefont{A.}~\bibnamefont{Buonanno}},
  \bibinfo{author}{\bibfnamefont{M.}~\bibnamefont{Lemoine}}, \bibnamefont{and}
  \bibinfo{author}{\bibfnamefont{K.~A.} \bibnamefont{Olive}},
  \bibinfo{journal}{Phys. Rev.} \textbf{\bibinfo{volume}{D62}},
  \bibinfo{pages}{083513} (\bibinfo{year}{2000}), \eprint{hep-th/0006054}.

\bibitem[{\citenamefont{Enqvist and Sloth}(2002)}]{Enqvist:2001zp}
\bibinfo{author}{\bibfnamefont{K.}~\bibnamefont{Enqvist}} \bibnamefont{and}
  \bibinfo{author}{\bibfnamefont{M.~S.} \bibnamefont{Sloth}},
  \bibinfo{journal}{Nucl. Phys.} \textbf{\bibinfo{volume}{B626}},
  \bibinfo{pages}{395} (\bibinfo{year}{2002}), \eprint{hep-ph/0109214}.

\bibitem[{\citenamefont{Lyth et~al.}(2003)\citenamefont{Lyth, Ungarelli, and
  Wands}}]{Lyth:2002my}
\bibinfo{author}{\bibfnamefont{D.~H.} \bibnamefont{Lyth}},
  \bibinfo{author}{\bibfnamefont{C.}~\bibnamefont{Ungarelli}},
  \bibnamefont{and} \bibinfo{author}{\bibfnamefont{D.}~\bibnamefont{Wands}},
  \bibinfo{journal}{Phys. Rev.} \textbf{\bibinfo{volume}{D67}},
  \bibinfo{pages}{023503} (\bibinfo{year}{2003}), \eprint{astro-ph/0208055}.

\bibitem[{\citenamefont{Moroi and Takahashi}(2001)}]{Moroi:2001ct}
\bibinfo{author}{\bibfnamefont{T.}~\bibnamefont{Moroi}} \bibnamefont{and}
  \bibinfo{author}{\bibfnamefont{T.}~\bibnamefont{Takahashi}},
  \bibinfo{journal}{Phys. Lett.} \textbf{\bibinfo{volume}{B522}},
  \bibinfo{pages}{215} (\bibinfo{year}{2001}), \eprint{hep-ph/0110096}.

\bibitem[{\citenamefont{Moroi and Takahashi}(2002)}]{Moroi:2002rd}
\bibinfo{author}{\bibfnamefont{T.}~\bibnamefont{Moroi}} \bibnamefont{and}
  \bibinfo{author}{\bibfnamefont{T.}~\bibnamefont{Takahashi}},
  \bibinfo{journal}{Phys. Rev.} \textbf{\bibinfo{volume}{D66}},
  \bibinfo{pages}{063501} (\bibinfo{year}{2002}), \eprint{hep-ph/0206026}.

\bibitem[{\citenamefont{Linde and Mukhanov}(1997)}]{Linde:1996gt}
\bibinfo{author}{\bibfnamefont{A.~D.} \bibnamefont{Linde}} \bibnamefont{and}
  \bibinfo{author}{\bibfnamefont{V.~F.} \bibnamefont{Mukhanov}},
  \bibinfo{journal}{Phys. Rev.} \textbf{\bibinfo{volume}{D56}},
  \bibinfo{pages}{535} (\bibinfo{year}{1997}), \eprint{astro-ph/9610219}.

\bibitem[{\citenamefont{Lyth and Wands}(2003{\natexlab{a}})}]{Lyth:2003ip}
\bibinfo{author}{\bibfnamefont{D.~H.} \bibnamefont{Lyth}} \bibnamefont{and}
  \bibinfo{author}{\bibfnamefont{D.}~\bibnamefont{Wands}},
  \bibinfo{journal}{Phys. Rev.} \textbf{\bibinfo{volume}{D68}},
  \bibinfo{pages}{103516} (\bibinfo{year}{2003}{\natexlab{a}}),
  \eprint{astro-ph/0306500}.

\bibitem[{\citenamefont{Stompor et~al.}(1996)\citenamefont{Stompor, Banday, and
  Gorski}}]{Stompor:1995py}
\bibinfo{author}{\bibfnamefont{R.}~\bibnamefont{Stompor}},
  \bibinfo{author}{\bibfnamefont{A.~J.} \bibnamefont{Banday}},
  \bibnamefont{and} \bibinfo{author}{\bibfnamefont{M.}~\bibnamefont{Gorski},
  \bibfnamefont{Krzysztof}}, \bibinfo{journal}{Astrophys. J.}
  \textbf{\bibinfo{volume}{463}}, \bibinfo{pages}{8} (\bibinfo{year}{1996}),
  \eprint{astro-ph/9511087}.

\bibitem[{\citenamefont{Enqvist et~al.}(2000)\citenamefont{Enqvist,
  Kurki-Suonio, and Valiviita}}]{Enqvist:2000hp}
\bibinfo{author}{\bibfnamefont{K.}~\bibnamefont{Enqvist}},
  \bibinfo{author}{\bibfnamefont{H.}~\bibnamefont{Kurki-Suonio}},
  \bibnamefont{and}
  \bibinfo{author}{\bibfnamefont{J.}~\bibnamefont{Valiviita}},
  \bibinfo{journal}{Phys. Rev.} \textbf{\bibinfo{volume}{D62}},
  \bibinfo{pages}{103003} (\bibinfo{year}{2000}), \eprint{astro-ph/0006429}.

\bibitem[{\citenamefont{Amendola et~al.}(2002)\citenamefont{Amendola, Gordon,
  Wands, and Sasaki}}]{Amendola:2001ni}
\bibinfo{author}{\bibfnamefont{L.}~\bibnamefont{Amendola}},
  \bibinfo{author}{\bibfnamefont{C.}~\bibnamefont{Gordon}},
  \bibinfo{author}{\bibfnamefont{D.}~\bibnamefont{Wands}}, \bibnamefont{and}
  \bibinfo{author}{\bibfnamefont{M.}~\bibnamefont{Sasaki}},
  \bibinfo{journal}{Phys. Rev. Lett.} \textbf{\bibinfo{volume}{88}},
  \bibinfo{pages}{211302} (\bibinfo{year}{2002}), \eprint{astro-ph/0107089}.

\bibitem[{\citenamefont{Crotty et~al.}(2003)\citenamefont{Crotty,
  Garcia-Bellido, Lesgourgues, and Riazuelo}}]{Crotty:2003rz}
\bibinfo{author}{\bibfnamefont{P.}~\bibnamefont{Crotty}},
  \bibinfo{author}{\bibfnamefont{J.}~\bibnamefont{Garcia-Bellido}},
  \bibinfo{author}{\bibfnamefont{J.}~\bibnamefont{Lesgourgues}},
  \bibnamefont{and} \bibinfo{author}{\bibfnamefont{A.}~\bibnamefont{Riazuelo}},
  \bibinfo{journal}{Phys. Rev. Lett.} \textbf{\bibinfo{volume}{91}},
  \bibinfo{pages}{171301} (\bibinfo{year}{2003}), \eprint{astro-ph/0306286}.

\bibitem[{\citenamefont{Gordon and Malik}(2004)}]{Gordon:2003hw}
\bibinfo{author}{\bibfnamefont{C.}~\bibnamefont{Gordon}} \bibnamefont{and}
  \bibinfo{author}{\bibfnamefont{K.~A.} \bibnamefont{Malik}},
  \bibinfo{journal}{Phys. Rev.} \textbf{\bibinfo{volume}{D69}},
  \bibinfo{pages}{063508} (\bibinfo{year}{2004}), \eprint{astro-ph/0311102}.

\bibitem[{\citenamefont{Beltran et~al.}(2004)\citenamefont{Beltran,
  Garcia-Bellido, Lesgourgues, and Riazuelo}}]{Beltran:2004uv}
\bibinfo{author}{\bibfnamefont{M.}~\bibnamefont{Beltran}},
  \bibinfo{author}{\bibfnamefont{J.}~\bibnamefont{Garcia-Bellido}},
  \bibinfo{author}{\bibfnamefont{J.}~\bibnamefont{Lesgourgues}},
  \bibnamefont{and} \bibinfo{author}{\bibfnamefont{A.}~\bibnamefont{Riazuelo}},
  \bibinfo{journal}{Phys. Rev.} \textbf{\bibinfo{volume}{D70}},
  \bibinfo{pages}{103530} (\bibinfo{year}{2004}), \eprint{astro-ph/0409326}.

\bibitem[{\citenamefont{Moodley et~al.}(2004)\citenamefont{Moodley, Bucher,
  Dunkley, Ferreira, and Skordis}}]{Moodley:2004nz}
\bibinfo{author}{\bibfnamefont{K.}~\bibnamefont{Moodley}},
  \bibinfo{author}{\bibfnamefont{M.}~\bibnamefont{Bucher}},
  \bibinfo{author}{\bibfnamefont{J.}~\bibnamefont{Dunkley}},
  \bibinfo{author}{\bibfnamefont{P.~G.} \bibnamefont{Ferreira}},
  \bibnamefont{and} \bibinfo{author}{\bibfnamefont{C.}~\bibnamefont{Skordis}},
  \bibinfo{journal}{Phys. Rev.} \textbf{\bibinfo{volume}{D70}},
  \bibinfo{pages}{103520} (\bibinfo{year}{2004}), \eprint{astro-ph/0407304}.

\bibitem[{\citenamefont{Kurki-Suonio et~al.}(2005)\citenamefont{Kurki-Suonio,
  Muhonen, and Valiviita}}]{KurkiSuonio:2004mn}
\bibinfo{author}{\bibfnamefont{H.}~\bibnamefont{Kurki-Suonio}},
  \bibinfo{author}{\bibfnamefont{V.}~\bibnamefont{Muhonen}}, \bibnamefont{and}
  \bibinfo{author}{\bibfnamefont{J.}~\bibnamefont{Valiviita}},
  \bibinfo{journal}{Phys. Rev.} \textbf{\bibinfo{volume}{D71}},
  \bibinfo{pages}{063005} (\bibinfo{year}{2005}), \eprint{astro-ph/0412439}.

\bibitem[{\citenamefont{Beltran et~al.}(2005)\citenamefont{Beltran,
  Garcia-Bellido, Lesgourgues, and Viel}}]{Beltran:2005gr}
\bibinfo{author}{\bibfnamefont{M.}~\bibnamefont{Beltran}},
  \bibinfo{author}{\bibfnamefont{J.}~\bibnamefont{Garcia-Bellido}},
  \bibinfo{author}{\bibfnamefont{J.}~\bibnamefont{Lesgourgues}},
  \bibnamefont{and} \bibinfo{author}{\bibfnamefont{M.}~\bibnamefont{Viel}},
  \bibinfo{journal}{Phys. Rev.} \textbf{\bibinfo{volume}{D72}},
  \bibinfo{pages}{103515} (\bibinfo{year}{2005}), \eprint{astro-ph/0509209}.

\bibitem[{\citenamefont{Bucher et~al.}(2004)\citenamefont{Bucher, Dunkley,
  Ferreira, Moodley, and Skordis}}]{Bucher:2004an}
\bibinfo{author}{\bibfnamefont{M.}~\bibnamefont{Bucher}},
  \bibinfo{author}{\bibfnamefont{J.}~\bibnamefont{Dunkley}},
  \bibinfo{author}{\bibfnamefont{P.~G.} \bibnamefont{Ferreira}},
  \bibinfo{author}{\bibfnamefont{K.}~\bibnamefont{Moodley}}, \bibnamefont{and}
  \bibinfo{author}{\bibfnamefont{C.}~\bibnamefont{Skordis}},
  \bibinfo{journal}{Phys. Rev. Lett.} \textbf{\bibinfo{volume}{93}},
  \bibinfo{pages}{081301} (\bibinfo{year}{2004}), \eprint{astro-ph/0401417}.

\bibitem[{\citenamefont{Seljak et~al.}(2006)\citenamefont{Seljak, Slosar, and
  McDonald}}]{Seljak:2006bg}
\bibinfo{author}{\bibfnamefont{U.}~\bibnamefont{Seljak}},
  \bibinfo{author}{\bibfnamefont{A.}~\bibnamefont{Slosar}}, \bibnamefont{and}
  \bibinfo{author}{\bibfnamefont{P.}~\bibnamefont{McDonald}},
  \bibinfo{journal}{JCAP} \textbf{\bibinfo{volume}{0610}}, \bibinfo{pages}{014}
  (\bibinfo{year}{2006}), \eprint{astro-ph/0604335}.

\bibitem[{\citenamefont{Bean et~al.}(2006)\citenamefont{Bean, Dunkley, and
  Pierpaoli}}]{Bean:2006qz}
\bibinfo{author}{\bibfnamefont{R.}~\bibnamefont{Bean}},
  \bibinfo{author}{\bibfnamefont{J.}~\bibnamefont{Dunkley}}, \bibnamefont{and}
  \bibinfo{author}{\bibfnamefont{E.}~\bibnamefont{Pierpaoli}},
  \bibinfo{journal}{Phys. Rev.} \textbf{\bibinfo{volume}{D74}},
  \bibinfo{pages}{063503} (\bibinfo{year}{2006}), \eprint{astro-ph/0606685}.

\bibitem[{\citenamefont{Trotta}(2007)}]{Trotta:2006ww}
\bibinfo{author}{\bibfnamefont{R.}~\bibnamefont{Trotta}},
  \bibinfo{journal}{Mon. Not. Roy. Astron. Soc. Lett.}
  \textbf{\bibinfo{volume}{375}}, \bibinfo{pages}{L26} (\bibinfo{year}{2007}),
  \eprint{astro-ph/0608116}.

\bibitem[{\citenamefont{Gupta et~al.}(2004)\citenamefont{Gupta, Malik, and
  Wands}}]{Gupta:2003jc}
\bibinfo{author}{\bibfnamefont{S.}~\bibnamefont{Gupta}},
  \bibinfo{author}{\bibfnamefont{K.~A.} \bibnamefont{Malik}}, \bibnamefont{and}
  \bibinfo{author}{\bibfnamefont{D.}~\bibnamefont{Wands}},
  \bibinfo{journal}{Phys. Rev.} \textbf{\bibinfo{volume}{D69}},
  \bibinfo{pages}{063513} (\bibinfo{year}{2004}), \eprint{astro-ph/0311562}.

\bibitem[{\citenamefont{Ferrer et~al.}(2004)\citenamefont{Ferrer, Rasanen, and
  Valiviita}}]{Ferrer:2004nv}
\bibinfo{author}{\bibfnamefont{F.}~\bibnamefont{Ferrer}},
  \bibinfo{author}{\bibfnamefont{S.}~\bibnamefont{Rasanen}}, \bibnamefont{and}
  \bibinfo{author}{\bibfnamefont{J.}~\bibnamefont{Valiviita}},
  \bibinfo{journal}{JCAP} \textbf{\bibinfo{volume}{0410}}, \bibinfo{pages}{010}
  (\bibinfo{year}{2004}), \eprint{astro-ph/0407300}.

\bibitem[{\citenamefont{Lemoine and Martin}(2007)}]{Lemoine:2006sc}
\bibinfo{author}{\bibfnamefont{M.}~\bibnamefont{Lemoine}} \bibnamefont{and}
  \bibinfo{author}{\bibfnamefont{J.}~\bibnamefont{Martin}},
  \bibinfo{journal}{Phys. Rev.} \textbf{\bibinfo{volume}{D75}},
  \bibinfo{pages}{063504} (\bibinfo{year}{2007}), \eprint{astro-ph/0611948}.

\bibitem[{\citenamefont{Multamaki et~al.}(2007)\citenamefont{Multamaki, Sainio,
  and Vilja}}]{Multamaki:2007hv}
\bibinfo{author}{\bibfnamefont{T.}~\bibnamefont{Multamaki}},
  \bibinfo{author}{\bibfnamefont{J.}~\bibnamefont{Sainio}}, \bibnamefont{and}
  \bibinfo{author}{\bibfnamefont{I.}~\bibnamefont{Vilja}}
  (\bibinfo{year}{2007}), \eprint{0710.0282}.

\bibitem[{\citenamefont{Kawasaki et~al.}(2005)\citenamefont{Kawasaki, Kohri,
  and Moroi}}]{Kawasaki:2004qu}
\bibinfo{author}{\bibfnamefont{M.}~\bibnamefont{Kawasaki}},
  \bibinfo{author}{\bibfnamefont{K.}~\bibnamefont{Kohri}}, \bibnamefont{and}
  \bibinfo{author}{\bibfnamefont{T.}~\bibnamefont{Moroi}},
  \bibinfo{journal}{Phys. Rev.} \textbf{\bibinfo{volume}{D71}},
  \bibinfo{pages}{083502} (\bibinfo{year}{2005}), \eprint{astro-ph/0408426}.

\bibitem[{\citenamefont{Moroi and Randall}(2000)}]{Moroi:1999zb}
\bibinfo{author}{\bibfnamefont{T.}~\bibnamefont{Moroi}} \bibnamefont{and}
  \bibinfo{author}{\bibfnamefont{L.}~\bibnamefont{Randall}},
  \bibinfo{journal}{Nucl. Phys.} \textbf{\bibinfo{volume}{B570}},
  \bibinfo{pages}{455} (\bibinfo{year}{2000}), \eprint{hep-ph/9906527}.

\bibitem[{\citenamefont{Spergel et~al.}(2007)}]{Spergel:2006hy}
\bibinfo{author}{\bibfnamefont{D.~N.} \bibnamefont{Spergel}}
  \bibnamefont{et~al.} (\bibinfo{collaboration}{WMAP}),
  \bibinfo{journal}{Astrophys. J. Suppl.} \textbf{\bibinfo{volume}{170}},
  \bibinfo{pages}{377} (\bibinfo{year}{2007}), \eprint{astro-ph/0603449}.

\bibitem[{\citenamefont{Lyth}(1985)}]{Lyth:1984gv}
\bibinfo{author}{\bibfnamefont{D.~H.} \bibnamefont{Lyth}},
  \bibinfo{journal}{Phys. Rev.} \textbf{\bibinfo{volume}{D31}},
  \bibinfo{pages}{1792} (\bibinfo{year}{1985}).

\bibitem[{\citenamefont{Martin and Schwarz}(1998)}]{Martin:1997zd}
\bibinfo{author}{\bibfnamefont{J.}~\bibnamefont{Martin}} \bibnamefont{and}
  \bibinfo{author}{\bibfnamefont{D.~J.} \bibnamefont{Schwarz}},
  \bibinfo{journal}{Phys. Rev.} \textbf{\bibinfo{volume}{D57}},
  \bibinfo{pages}{3302} (\bibinfo{year}{1998}), \eprint{gr-qc/9704049}.

\bibitem[{\citenamefont{Lyth and Wands}(2003{\natexlab{b}})}]{Lyth:2003im}
\bibinfo{author}{\bibfnamefont{D.~H.} \bibnamefont{Lyth}} \bibnamefont{and}
  \bibinfo{author}{\bibfnamefont{D.}~\bibnamefont{Wands}},
  \bibinfo{journal}{Phys. Rev.} \textbf{\bibinfo{volume}{D68}},
  \bibinfo{pages}{103515} (\bibinfo{year}{2003}{\natexlab{b}}),
  \eprint{astro-ph/0306498}.

\bibitem[{\citenamefont{Weinberg}(2004)}]{Weinberg:2004kf}
\bibinfo{author}{\bibfnamefont{S.}~\bibnamefont{Weinberg}},
  \bibinfo{journal}{Phys. Rev.} \textbf{\bibinfo{volume}{D70}},
  \bibinfo{pages}{083522} (\bibinfo{year}{2004}), \eprint{astro-ph/0405397}.

\end{thebibliography}

\end{document}